\documentstyle[12pt]{article}
\begin{document}
\vskip 2 cm
\begin{center}
\large{\bf
A " RIP VAN WINKLE HYPOTHESIS " TO RESOLVE THE
REALISM - ANTIREALISM DEBATE}
\vskip 3 cm
{\bf Afsar Abbas}\\
Institute of Physics, Bhubaneswar - 751005, India\\
{afsar@iopb.res.in}
\vskip 4 cm
{\bf ABSTRACT}
\end{center}
\vskip 2 cm

The intensity of debate between the realists and antirealists 
shows no sign of abating. Here a new hypothesis is proposed
to resolve the issue. The requirement of consistency and 
continuity are built-in in the methodology of this hypothesis.
This new hypothesis supports realism.

\newpage

A persistent feature in the field of philosophy 
of science is that of antagonism between the respective advocates
of realism and anti-realism. The debate has continued since
antiquity and shows no sign of petering out. Both the sides have
diligently stuck to their guns. In the present context,
Nancey Murphy (1990) summarizes, ".. there are both modern and
postmodern versions of realism and antirealism - hence the
confusion, the inability to be quite sure of what one is 
arguing against". 

This discussion between the various adherents of realism and
those of antirealism is well documented in 
books and treatises on 
philosophy of science. 
I do not intend to go into details of the same here 
and would refer the reader 
to a clear and neat exposition of the issues involved
by James Ladyman ( Ladyman (2002) ).

If this is the state of affairs, then clearly
the question that one has to address is whether we have been 
missing some essential aspect in this debate between the 
realists and the antirealists. 
Given the significance of the issues
involved, any new framework, which would help in clarifying 
the discussion should be welcome.
Here I would like to present a
new perspective to resolve this debate. This is done by
proposing a new "Rip Van Winkle hypothesis".

To understand it, let me first remind the reader of 
the story of Rip Van
Winkle. Washington Irving (1783 - 1859) wrote "Rip Van Winkle and
the legend of sleepy hollow". The story is that of a Dutch
by the name of Rip Van Winkle. He was an easy going fellow
who liked to play with kids and did not want be bothered 
with work. He had a nagging wife. 
To escape her nagging he would often go wandering in the nearby
Catskill Mountains with his faithful dog named Wolf. In one 
such trip he came across a group of 
dwarfs who were playing ninepins. 
He drank some brew from a keg they had with them and as a
result of this he went to sleep. When he woke up he found 
that his dog was gone and he also saw a rusted
weapon rather than his shiny rifle, lying by his side. 
He was extremely puzzled. On coming back
to his village, he came to the amazing realization that he had
woken up after some 20 years of sleep.

Let us conduct a Gedanken experiment ( Gedanken in German 
means "thought" ). Such gedanken experiments have been 
extensively used in science and in  
physics in particular to reach consistent and
"correct" conclusions. Gedanken experiment is a powerful 
tool in science.

Imagine two persons named respectively "Antirealist" and
"Realist". Mr. Antirealist is a 
hardcore antirealist living in 17th Century 
London. Mr. Realist is a committed realist who loves to argue
with Mr. Antirealist. ( Here we are using the words
realism and antirealism in the canonical manner that these
words are used today in the field of philosophy of science
in as broad a context as possible ( Ladyman ( 2002) ) ). 
Once they both went wandering into nearby woods. 
They came across some strange
characters who had a cask of tempting brew with them. 
Being thirsty they could not resist drinking it
and immediately went to
sleep, just like what Rip Van Winkle had done. 
But these two slept for a much longer period 
and let us say, they wake up today in the 21st Century. How will
they react to the new realities that they shall encounter?     

Mr. Antirealist will be completely flabbergasted.
His own antirealistic
philosophical convictions would not allow him to make sense of
what he shall observe. He would prefer to think that he is
dreaming. He would have to make arbitrary assumptions like 
supernatural powers, magic etc to make sense of
what he would be confronted with.
 
Mr. Realist, after overcoming the initial shock, 
would smile and say, " Aha, I understand it all. 
Just like in gravity and magnetism one
has fields which exert forces far away, I can understand
transmission of sound and pictures through 
some advanced technology
as is available today. Well, that is your TV and mobile phone. 
I always fancied myself flying like a bird. They, through
ingenious use of science have made these aeroplanes. The light
bulbs are nothing more than controlled energy. What I
could only imagine in my wildest dreams has actually come 
to pass. Wonderful! Quite clearly I was correct in 
believing in the reality of science and my friend Antirealist was
quite clearly off the track. I hope he realizes that now."

From this gedanken "experiment" 
one may conclude that Mr. Realist was correct
because his philosophy and understanding of nature was shown
to be consistent and continuous. That is - 
consistency and continuity of
his philosophical perspective over a span of several hundred years
confirms the veracity of his ideology.
On the other hand Mr Antirealist
had to make inconsistent and sudden breaks in his perspective of
the physical world. Self consistency and continuity are essential
minimum requirements for the correctness  
of an ideological framework.    

Existence of consistency is an extremely significant requirement
of any theoretical or mathematical framework in physics or 
mathematics.
In the 1920's the distinguished mathematician D. Hilbert
( Hilbert (1926) ) launched an ambitious programme to provide a
formalization of both logic and arithmetic and which would prove
rigorously the complete consistency of the fundamental axioms. 
No one doubted that mathematics being so basic should
indeed be self-consistent as well.
However Kurt Goedel shocked everyone by showing that Hilbert's
expectations were absolutely wrong! He showed ( Goedel (1931) )
that finitary proofs of self-consistency cannot be established
within the formalism of classical mathematics, or of set theory,
or of the axioms of Russell and Whitehead's Principia Mathematica
supplemented by Peano's axioms.
This was a shocking blow to the consistency programme in
mathematics. This shows
how significant can the requirement of consistency be for any
discipline of knowledge. 

The gedanken experiment exemplified by 
the Rip Van Winkle hypothesis above proves that
due to the in-built consistency, the hypothesis of realism is correct. 
Note that here self-consistency means that what was empirically
and theoretically understood to be true in the 
17th century continues to be so at present in the 21st century.
Science would have grown in the intervening period, and so 
it requires that there be just simple 
and natural embedding of the physical structure
known then, into the present physical framework.
On the other, hand antirealism fails miserably 
in this consistency test.

Equally significant is the test of continuity which is implied by
the Rip Van Winkle hypothesis. This means that
an ideology which can be 
demonstrated to hold continuously over a long period of time 
( for example in this case for several hundred years ), 
demonstrates its resilience and intrinsic strength which can
necessarily arise from it being on the right track.
Realism has it and antirealism does not.

Note that the Rip Van Winkle hypothesis propounded here in support
of realism is very different from Putnam's "no-miracles"
hypothesis, which supports realism by pointing out that it is
the only philosophy which does not require that the success of
science be treated as a miracle. 
The no-miracles hypothesis directly compares realism and
antirealism at a particular instant of time. So basically it is a 
"static" test of realism. 
By static one means that there is no 
concept of flow of time herein. 
On the other hand the Rip Van Winkle 
hypothesis is a dynamic test of realism. Dynamic means that 
realism still holds true in spite of passage of time.
In addition the Rip Van Winkle hypothesis gains strength
by relying upon the requirement of "consistency" and
"continuity" to confirm the veracity of the ideology of realism. 
These ideas do not 
figure in the no-miracles hypothesis, which relies directly
upon the concept of miracle itself.

\newpage

\vskip 2 cm
{\bf REFERENCES}
\vskip 1 cm

Goedel, K. (1931), " Ueber formal untenstche ... ",
Mh. Math. Phys., 38, 173-198

Hilbert, D. (1926), " Ueber das unendliche ", Math. Annln, 95 ,
161-190 

Ladyman, J. (2002), " Understanding philosophy of science ",
Routledge, London

Murphy, N. (1990), " Scientific realism and postmodern
philosophy ". Brit. J. Phil. Sci., 41, 291-303

Putnam, H. (1975), " Mathematics, matter and method: philosophical
papers " Vol I, Cambridge University Press, Cambridge

\end{document}